\documentclass[aps,prl,twocolumn,floatfix, preprintnumbers]{revtex4}
\usepackage[pdftex]{graphicx}
\usepackage[mathscr]{eucal}
\usepackage[pdftex]{hyperref}
\usepackage{amsmath,amssymb,bm}

\newif\ifarxiv
\arxivfalse

\begin		{document}

\def\M		{M}

\def\section	#1{\quad\textit{#1}.---}
\def\suck[#1]#2{\includegraphics[#1]{#2}}        


\title
    {
    Constraining Relativistic Generalizations of Modified Newtonian Dynamics with Gravitational Waves
    }

\author{Paul~M.~Chesler}
\email{pchesler@g.harvard.edu}

\author{Abraham Loeb}
\affiliation
    {Black Hole Initiative, Harvard University, Cambridge, MA 02138, USA}
\email{aloeb@cfa.harvard.edu}

\date{\today}

\begin{abstract}
In the weak-field limit of General Relativity, gravitational waves obey linear equations 
and propagate at the speed of light.  These properties of General Relativity are 
supported by the observation of ultra high energy cosmic rays as well as by LIGO's recent detection of gravitation waves.  We argue that two existing relativistic generalizations of Modified Newtonian 
Dynamics, namely Generalized Einstein-Aether theory and BIMOND, display fatal inconsistencies with these observations.
\end{abstract}


\maketitle
\parskip	2pt plus 1pt minus 1pt

\section{Introduction}With LIGO's  discovery of black hole mergers \cite{Abbott:2016blz,Abbott:2016nmj},
gravitational waves have advanced from a theoretical prediction of 
General Relativity (GR) to an experimental detection.  
Gravitational waves also exist in modified theories of gravity and can have distinguishing 
features which can be used to constrain the theoretical space of allowable theories.  
Of particular interest are relativistic generalizations of Modified Newtonian Dynamics (MOND) \cite{Milgrom:1983ca} 
 (henceforth referred to as MONDian theories).
These are modified theories of gravity designed to 
emulate the effects of dark matter without actually having dark matter (see \cite{Famaey:2011kh} for a review).  

At the core of MONDian theories is the assumption that in the limit of 
sufficiently weak acceleration, gravitational dynamics becomes non-linear, with the non-linearities 
tailored to yield flat rotation curves of galaxies %
\footnote
  {
  We note that there also exist formulations of MOND which modify interia instead of 
  gravity \cite{Milgrom:1992hr,Milgrom:2011kx}.  In this paper we only consider 
  modifications of gravity.
  }.  
In the weak-field quasi-static limit, where the metric can be written
$ds^2 = -(1 + 2 \Phi) dt^2 + (1 + 2 \Psi) d \bm x^2,$
the field equations must reduce to $\Psi = -\Phi$ and  \cite{Famaey:2011kh}  
\begin{align}
\label{eq:MondEqns}
\nabla \cdot [ \mu \left ( {\textstyle \frac{|\nabla \Phi|}{a_0}} \right ) \nabla \Phi ] = 4 \pi G \rho,&
\end{align}
where $G$ is Newton's constant, $a_0 \approx 10^{-10} \ {\rm m/s}^2$, and $\rho$ is the matter density.  
The  MOND function $\mu(x)$ satisfies $\mu(x) \to 1$ when $x \gg 1$, so that Newton's 
law of gravity is recovered in the strong acceleration limit, and $\mu(x) \to x$ when $x \ll 1$, which 
yields flat rotation curves of galaxies at large distances from matter sources.
Hence, the equations of motion for the potential become non-linear when $|\nabla \Phi| \ll a_0$.
This stands in stark contrast to GR,
where the weak-field limit is governed by linear equations of motion.   
MONDian theories of gravity include TeVeS \cite{Bekenstein:2004ne}, generalized Einstein-Aether theories
\cite{Zlosnik:2006zu}, bimetric theories (BIMOND) \cite{Milgrom:2009gv,Milgrom:2013iea},
and non-local theories \cite{Deffayet:2011sk,Deffayet:2014lba}.
Recently Verlinde \cite{Verlinde:2016toy} suggested that similar modifications
can naturally occur in entropic gravity \cite{Verlinde:2010hp}.  

MONDian modifications to GR can potentially alter 
gravitational wave physics in at least two ways.
First, since MOND is an acceleration based 
modification of gravity, MONDian theories can violate the equivalence principle.  
A consequence of this is that gravitational waves 
can propagate subluminally.  Second, since MONDian 
theories are non-linear in the weak field limit, gravitational 
waves can be governed by non-linear equations, even in the weak-field limit.
As we elaborate on below, these features have unsavory consequences and can be used to
restrict the set of allowed MONDian theories.

If gravitational waves propagate non-luminally, the arrival time for electromagnetic and gravitational 
signals from astrophysical events can be different \cite{Kahya:2007zy,Kahya:2010dk}.  Moreover,   
as was pointed out long ago \cite{Moore:2001bv},
if the speed of gravitational waves is $c_{\rm g} < 1$ (in units where $c = 1$), 
then high energy 
cosmic rays traveling at speed $v \to 1$ will lose energy via the emission of gravitational 
Cherenkov radiation, with an energy loss rate dependent on the difference $1 - c_{\rm g}$.
The observation of high energy cosmic rays on earth, combined with an estimate of their distance of 
propagation,
then sets lower bounds on $c_{\rm g}$, which have been estimated to be
$1 - c_{\rm g} \lesssim 10^{-15}$ \cite{Moore:2001bv,Elliott:2005va}. 
In the MOND limit of the Einstein-Aether theory of Ref.~\cite{Zlosnik:2006zu}, 
we demonstrate that the speed of gravitational waves depends on the local 
gravitational potential and generically cannot be set equal to the speed of light,
and that Cherenkov losses are unavoidable without making the theory pathological.
These features make this theory an unacceptable theory of gravity.  

Second, if 
gravitational wave dynamics are non-linear in the weak-field limit, 
gravitational waves emitted in black hole merger events can interact with themselves as well as 
with other gravitational waves, effectively scrambling the structure of the original waveforms 
as they propagate to earth.
LIGO's recent observation of GW150914 had a gravitational waveform 
completely consistent with GR \cite{Abbott:2016blz}, suggesting no such scrambling effect.  
A natural expectation is therefore that gravitational waves 
must satisfy linear equations of motion in the weak-field
limit of any acceptable theory of gravity.

We argue that 
interactions between gravitational wave packets 
in the weak-field limit of BIMOND \cite{Milgrom:2013iea} alters the structure of the original 
waveforms and can even lead to singular evolution.
Therefore, if BIMOND reduces to GR in the strong field limit --- and thereby yields
the same \textit{initial} gravitational waveforms as GR in merger events --- the  
waveforms observed far away would not look anything like those predicted by GR.
In BIMOND we argue non-linear interactions become important 
at distances on the order 0.3 Gpc from merger events.
In contrast, gravitational waves from
GW150914 are estimated to have propagated 0.4 Gpc.  Our results and the experimental data from LIGO suggest that BIMOND in its present form is not an acceptable  theory of modified gravity.

\section{Generalized Einstein-Aether Theories}%
In addition to the metric $g_{\mu \nu}$, Einstein-Aether theories contain 
a time-like vector field $A^\mu$ which satisfies $A^2 = -1$ and 
defines a preferred frame.  
Following Ref.~\cite{Zlosnik:2006zu} we consider the gravitational action,
\begin{equation}
\label{eq:AEaction}
S = \frac{1}{16 \pi G} \int d^4 x \sqrt{g} \left [ R + \M^2 
\mathcal F({\textstyle \frac{\mathcal K}{\M^2}}) + \lambda (A^2 {+} 1)\right  ]
+ S_{\rm mat},
\end{equation}
where $\mathcal K \equiv \mathcal K^{\alpha \beta}_{\ \ \gamma \sigma}
\nabla_\alpha A^\gamma \nabla_\beta A^\sigma$ is a quadratic function of derivatives of $A^\mu$,
 $\lambda$ is a 
Lagrange multiplier which enforces the constraint $A^2 = -1$, $M$ is a constant
with dimensions of mass, and $S_{\rm mat}$ is the matter action.
The  function $\mathcal F$ determines the 
function $\mu$ in (\ref{eq:MondEqns}).
The most general expression
for $\mathcal K^{\alpha\beta}_{\ \ \mu \nu}$ involving no derivatives reads
\begin{equation}
\mathcal K^{\alpha \beta}_{\ \ \gamma \sigma} \equiv c_1 g^{\alpha \beta} g_{\gamma \sigma}
+ c_2 \delta^{\alpha}_{\gamma} \delta^{\beta}_{\sigma}
+ c_3 \delta^{\alpha}_\sigma \delta^\beta_\gamma
+ c_4 A^\alpha A^\beta g_{\sigma \gamma},
\end{equation}
where the $c_i$ are dimensionless constants.
Following \cite{Zlosnik:2006zu} we shall set $\M = \epsilon^2 a_0$
with $\epsilon$ a bookkeeping parameter which can be set
to one after all calculations.

The Einstein-Aether equations of motion read
\begin{subequations}
\label{eq:AEeqm}
\begin{eqnarray}
R_{\mu \nu} - {\textstyle \frac{1}{2} }R  g_{\mu \nu} &=& \mathcal T_{\mu \nu} + 8 \pi G T_{\mu \nu}^{\rm mat},\\
\nabla_\alpha [ \mathcal F' J^\alpha_{\ \beta}] - \mathcal F' y_\beta &=& 2 \lambda A_{\beta},
\end{eqnarray}
\end{subequations}
with $T_{\mu \nu}^{\rm mat}$ the matter stress and $\mathcal T_{\alpha \beta}$ the vector stress, 
\begin{eqnarray}
\label{eq:vectorstress}
\mathcal T_{\alpha \beta} &=& {\textstyle \frac{1}{2}} 
\nabla_\sigma \{ \mathcal F' [J^{\ \ \sigma}_{(\alpha} A_{\beta)}
- J^\sigma_{\ ( \alpha} A_{\beta)} - J_{(\alpha \beta)} A^\sigma]\}
\\ \nonumber 
&-& \mathcal F' Y_{\alpha \beta} + {\textstyle \frac{1}{2}} g_{\alpha \beta} \M^2 \mathcal F
+\lambda A_{\alpha} A_{\beta},
\end{eqnarray}
with 
\begin{subequations}
\begin{eqnarray}
Y_{\alpha \beta} &=& - c_1 [ \nabla_\nu A_\alpha \nabla^\nu A_\beta 
- \nabla_\alpha A_\nu \nabla_\beta A^\nu ] \\ \nonumber
&&-c_4 (A \cdot \nabla A_\alpha) (A \cdot \nabla A_\beta), 
\\
J^\alpha_{\ \sigma} &=& 2 \mathcal K^{\alpha \beta}_{\ \ \sigma \gamma} \nabla_\beta A^\gamma,
\\
y_\beta &=& 2 c_4 \nabla_\beta A_\mu (A \cdot \nabla A^\mu). 
\end{eqnarray}
\end{subequations}

We wish to study Einstein-Aether waves in the MOND limit,
particularly waves propagating in the background of a weak, static and slowly varying
gravitational field.  Our goal here is to demonstrate that the propagation speeds 
depend on the local background
fields and cannot be set equal to the speed of light for all modes.
To this end let us first consider static, weak field, and slowly varying solutions to the Einstein-Aether
system.  Following \cite{Zlosnik:2006zu} we consider the ansatz,
\begin{subequations}
\begin{eqnarray}
g_{\mu \nu}(t,\bm x) &=& \eta_{\mu \nu} - 2 \epsilon \, \Phi(\epsilon \bm x) \delta_{\mu \nu}, 
\\
A_\mu(t,\bm x) &=& [-1 + \epsilon \, \Phi(\epsilon \bm x)] \delta_{\mu 0},
\end{eqnarray}
\end{subequations}
and solve the equations of motion in the  $\epsilon \to 0$ limit.
The above ansatz satisfies $A^2 = -1 +O(\epsilon^2)$.  
In the $\epsilon \to 0$ limit the Einstein-Aether equations of motion  (\ref{eq:AEeqm})
reduce to the MOND equation
(\ref{eq:MondEqns}) with $\mu(x) = x$ provided
\cite{Zlosnik:2006zu}
\begin{equation}
\label{eq:Fdef0}
\textstyle \mathcal F(x) = \frac{1}{-c_1+c_4} \left [2 x - \frac{4 }{3  \sqrt{-c_1+c_4}} x^{3/2} \right].
\end{equation}

We note that the Einstein-Aether theory
can be made GR-compatible in the strong field limit if one sets $\mathcal F(x) = \xi x$.  
By setting $\xi \to 0$, the vector stress (\ref{eq:vectorstress}) can be made arbitrarily small.  
Hence, constraints from strong field gravity, such as those studied in \cite{Bonetti:2015oda,Yagi:2013qpa,Yunes:2016jcc}, can be satisfied. 

Consider now the ansatz
\begin{subequations}
\label{eq:EApropansatz}
\begin{eqnarray}
g_{\mu \nu}(t,\bm x) &=& \eta_{\mu \nu} - 2 \epsilon \Phi(\epsilon \bm x) \delta_{\mu \nu} 
+  \zeta \, h_{\mu \nu} e^{- i \omega t + i \bm k \cdot \bm x}, \ \ \ \ \ 
\\
A_\mu(t,\bm x) &=& [-1 + \epsilon \Phi(\epsilon \bm x)] \delta_{\mu 0} 
+ \zeta \, a_\mu  e^{- i \omega t + i \bm k \cdot \bm x},
\end{eqnarray}
\end{subequations}
which describes small perturbations propagating on top of the 
static background potential $\Phi$.  
Here $\zeta$ is another bookkeeping parameter which
parameterizes the strength of the propagating modes.  We shall 
consider the $\epsilon \to 0$ limit with $\zeta \ll \epsilon^2$.  In this limit the exponentials 
vary in space much more rapidly than the potential.  Note that $h_{\mu \nu}$
and $a_{\mu}$ also depend on $\bm x$.  However, this dependence can
be neglected at leading order.  For simplicity we assume the potential vanishes
as the point $\bm x$ of interest 
and that $\bm k$ and $\nabla \Phi$ point in the same direction at $\bm x$.

The equations of motion in the MOND limit for $h_{\mu \nu}$ and $a_{\mu}$, as well as the dispersion
relation $\omega(k)$, follow from substituting the ansatz (\ref{eq:EApropansatz})
into (\ref{eq:AEeqm}) with $\mathcal F$ given by Eq.~(\ref{eq:Fdef0}).
With $\zeta \ll \epsilon^2$ and the presence of the background potential,
the equations of motion for $h_{\mu \nu}$ and $a_\mu$ 
are linear.  There are a total of five propagating modes, 
including two tensor 
modes, two vector modes, and one scalar mode.   We find linear dispersion relations 
$\omega = c_{\rm g} k$ for all modes, with propagation speeds,
\begin{widetext}
\begin{subequations}
\label{eq:speeds}
\begin{eqnarray}
c_{\rm tensor}^2 &=& \frac{-c_1+c_4}{(c_1 + 2 c_3+c_4) - 2 (c_1 + c_3) | \nabla \Phi|/a_0}, \\
c_{\rm vector}^2 &=& \frac{c_3^2 -c_1 c_4+ (c_1^2 - c_3^2) | \nabla \Phi|/a_0}
{(-c_1+c_4) [c_1 + 2 c_3+c_4  - 2 c_1 (c_1 + c_3) | \nabla \Phi|/a_0]},  \\
\label{eq:scalarvel}
c_{\rm scalar}^2 &=& \frac{2 (-c_1+c_4) (c_1 + c_2+c_3)(1 -| \nabla \Phi|/a_0)| \nabla \Phi|/a_0 }
{(1 -2 | \nabla \Phi|/a_0)[c_1 + 2 c_3 +c_4- 2 (c_1 + c_3)| \nabla \Phi|/a_0][2 c_1 + 3 c_2+c_3 -c_4 - (c_1+3 c_2 + c_3)| \nabla \Phi|/a_0]}. \ \ \ \ \ \ 
\end{eqnarray}
\end{subequations}
\end{widetext}
Note that the propagation speeds depend on the local value of the potential gradient, $|\nabla \Phi|/a_0$.

Setting the speed of all excitations equal to the speed of light will guarantee no cosmic ray energy loss
via gravitational Cherenkov radiation.  Let us focus on the tensor and scalar modes. 
Inspection of Eq.~(\ref{eq:speeds}) shows that setting $c_{\rm tensor} = 1$ 
requires $c_3 = -c_1$.  In contrast, no choice of $c_i$ can set $c_{\rm scalar} = 1$.  Nevertheless, the scalar mode can be made non-propagating, meaning $c_{\rm scalar} = 0$, if in addition to $c_3=-c_1$ we set $c_2 = 0$.  

There is another reason that one should set $c_3 = -c_1$.  As we shall discuss 
further below in the next section, in the absence of background potentials (\textit{i.e.} $\Phi = 0$)
it is desirable that gravitational waves should satisfy linear equations in the weak field limit.  For the tensor modes %
this can only be accomplished if
$c_3 = -c_1$ 
\footnote
  {  
  We note that even after setting $c_3 = -c_1$,  the scalar and vector modes
  still satisfy non-linear equations.
  }.

It is easy to see that the Einstein-Aether theory with $c_3 = -c_1$ and $c_2 = 0$ develops 
pathologies, even with regular initial data.  The pathologies are essentially the MONDian 
equivalent of those found in the Einstein-Aether theory studied in \cite{Jacobson:2000xp}.  
In particular, with regular initial
data the vector field will generically evolve to a singular solution.  To see this, first note
that with 
$c_3 = -c_1$ and $c_2 = 0$ we have,
\begin{equation}
\mathcal K = -\frac{c_1}{2} F_{\mu \nu} F^{\mu \nu} + c_4 (A \cdot \nabla A)^2,
\end{equation}
where $F_{\mu \nu} = \partial_\mu A_\nu - \partial_\nu A_\mu$ is the field
strength of $A$.  Now consider \textit{pure gauge} solutions   
in which $F_{\mu \nu} = 0$. From $F_{\mu \nu} = 0$, it follows that
$\nabla_\alpha A^2 =  2 A \cdot \nabla A_\alpha = 0$, meaning 
that $A$ is tangent to a congruence of time-like geodesics.  Such vector fields
have $\mathcal K = 0$, $J_{\mu \nu} = Y_{\mu \nu} = 0$, 
and are exact solutions to the equation of motion  (\ref{eq:AEeqm}).
They also have vanishing stress $\mathcal T^{\mu \nu}$, meaning they do not 
back react on the geometry.  Consider then a solution to the Einstein equations 
sourced by some matter distribution.  Generically the congruence of geodesics
will form caustics.  When this happens 
$\nabla_\alpha A_\beta$ will become singular, and the classical theory will break down.
Therefore, demanding no Cherenkov energy losses --- by setting $c_3 = -c_1$ and $c_2 = 0$ --- results in a non-viable theory.

\section{BIMOND}%
In addition to the physical metric $g_{\mu \nu}$, BIMOND \cite{Milgrom:2009gv,Milgrom:2013iea} contains an additional 
metric $\hat g_{\mu \nu}$.  The difference between the Christoffel symbols associated
with the two metrics,
\begin{equation}
C^\lambda_{\mu \nu} = \Gamma^\lambda_{\mu \nu} - \hat \Gamma^\lambda_{\mu \nu},
\end{equation}
is itself a tensor.  We will refer to $C^\lambda_{\mu \nu}$ as the acceleration tensor.  The 
MOND limit is defined by $|C^\lambda_{\mu \nu}| \lesssim a_0$.  
In the limit $|C^\lambda_{\mu \nu}| \gg a_0$, interactions between the two metrics
are assumed to vanish and the dynamics of $g_{\mu \nu}$ are governed by GR.

Before writing down the action of BIMOND, let us first estimate the magnitude of the 
acceleration tensor associated with gravitational waves emitted by merging black holes 
at a distance $r$ away from an observer.  In the weak-field limit, the acceleration tensor 
scales as $h /\lambda$ where 
$h$ is the amplitude of the wave and $\lambda$ its wavelength.  For equal mass binary black hole mergers the 
amplitude roughly scales like $h \sim G M/r $, with $M$ being the mass of
the binary.  Likewise, for equal mass mergers the wavelength of emitted
radiation should be set
by the Schwarzschild radius, implying that $\lambda \sim G M$.  We therefore obtain,
\begin{equation}
|C^\lambda_{\mu \nu}| \sim \frac{1}{r}.
\end{equation}
This is equal to $a_0$ when $r \sim 0.3$ Gpc.  This suggests that gravitational waves 
from GW150914, which are believed to have originated 0.4 Gpc away, have probed the 
MONDian limit.  Indeed, for GW150914 the \textit{observed} amplitude of the gravitational
waves was $h \sim 10^{-21}$ while the wavelength was $\lambda \sim 5 \times 10^6$ m, which yields
$h/\lambda \approx 0.2 a_0$.
Moreover, it is likely that these gravitational waves propagated 
through cosmic voids, where the acceleration tensor due to quasi-static matter sources is  
less than the acceleration produced by the gravitational waves.  We therefore chose
to study gravitational waves propagating in the absence of background potentials.
The equations of motion for such gravitational waves in BIMOND were derived in Ref.~\cite{Milgrom:2013iea}, as we briefly sketch below.

Following Milgrom \cite{Milgrom:2013iea}, we consider the 
gravitational action
\begin{equation}
\label{eq:GravAction}
S_{\rm grav} = \frac{1}{16 \pi G} \int d^4 x \left [ \alpha \sqrt{g} R + \beta \sqrt{\hat g} \hat R 
+ M^2 \mathcal F \right ],
\end{equation} 
where again $M = \epsilon a_0$ with $\epsilon$ a bookkeeping parameter.
Here $R$ and $\hat R$ are the Ricci scalars of $g_{\mu \nu}$ and $\hat g_{\mu \nu}$, respectively,
and $\alpha$ and $\beta$ are constants.  We shall consider the 
case $\alpha \neq -\beta$.  $\mathcal F$ characterizes the interactions
between the two metrics.  Derivative interactions between $g_{\mu \nu}$
and $\hat g_{\mu \nu}$ are assumed to solely be contained in the dimensionless scalar arguments 
$ \frac{1}{M^2} g^{\mu \nu} \Upsilon_{\mu \nu}$
and $ \frac{1}{M^2}  \hat g^{\mu \nu} \Upsilon_{\mu \nu}$ with
\begin{align}
\Upsilon_{\mu \nu} \equiv C^\gamma_{\mu \lambda} C^\lambda_{\nu \gamma}-C^\gamma_{\mu \nu} C^\lambda_{\lambda \gamma}.
\end{align}
Note that additionally $\mathcal F$ 
can depend on the non-derivative interactions such as
$g \hat g$ or  $g^{\mu \nu} \hat g_{\mu \nu}$.  

In the weak field  limit we can write 
\begin{align}
&g_{\mu \nu} = \eta_{\mu \nu} +\epsilon  h_{\mu \nu},&
\hat g_{\mu \nu} = \eta_{\mu \nu} + \epsilon \hat h_{\mu \nu}.&
\end{align}
We wish to obtain the equations of motion for $h_{\mu \nu}$ and 
$\hat h_{\mu \nu}$ at first order in $\epsilon$.  
At first order any dependence on non-derivative 
interactions between $h_{\mu \nu}$ and $\hat h_{\mu \nu}$ can be neglected.  
Likewise, in this limit 
$ \frac{1}{M^2} g^{\mu \nu} \Upsilon_{\mu \nu} = \frac{1}{M^2}  \hat g^{\mu \nu} \Upsilon_{\mu \nu} \equiv \frac{1}{M^2}\Upsilon$
where, 
\begin{equation}
\Upsilon \equiv \eta^{\mu \nu} \Upsilon_{\mu \nu}.
\end{equation}
Therefore, in the weak field  limit we can take,
\begin{equation}
\label{eq:Fdef}
\mathcal F = \mathcal F(-\Upsilon/2 M^2).
\end{equation}
For quasi-static geometries (and $\alpha \neq -\beta$) MOND phenomenology is obtained when
\cite{Milgrom:2013iea}
\begin{align}
\label{eq:scalingrelation}
&\mathcal F'(x) + \frac{\alpha \beta}{\alpha + \beta}  \sim \sqrt{x},& \ \ x \ll 1.&
\end{align}
For GW150914 we estimate $\Upsilon \sim 0.04 \, a_0^2$.
Hence we employ (\ref{eq:scalingrelation}) for studying gravitational waves.

The equations of motion for the metric perturbations can easily be
derived from the action (\ref{eq:GravAction}) and Eqs.~(\ref{eq:Fdef})
and (\ref{eq:scalingrelation}).
Define, 
\begin{align}
&\Delta h_{\mu \nu} \equiv h_{\mu \nu} - \hat h_{\mu \nu},&
s_{\mu \nu} \equiv \alpha h_{\mu \nu} + \beta \hat h_{\mu \nu}.&
\end{align}
Since $\mathcal F$ only depends on $\Delta h_{\mu \nu}$, 
the weak field action for $s_{\mu \nu}$ is simply that of linearized
Einstein equations, meaning $s_{\mu \nu}$ satisfies the linearized
Einstein equations.  In contrast, up to inconsequential surface terms 
the weak field action for $\Delta h_{\mu \nu}$
is \cite{Milgrom:2013iea},
\begin{equation}
\label{eq:nonlinearaction}
S \propto \int d^4 x \, (-\Upsilon)^{3/2}.
\end{equation}
It follows that $\Delta h_{\mu \nu}$ is governed by non-linear equations. 

To illustrate the non-linear nature of the equations of motion and the resulting
instabilities they possess, let us focus on waves propagating in the $z$ direction
with a single excited mode $\Delta h_{xy}(t,z)$, with $x,y$ the two transverse directions
to $z$.  In this case, the action (\ref{eq:nonlinearaction}) leads to the equation of motion,
\begin{equation}
\label{eq:hxyeq}
\partial_\mu \{ | \partial \Delta h_{xy}| \partial^\mu \Delta h_{xy} \} = 0,
\end{equation}
where $| \partial \Delta h_{xy}| \equiv \sqrt{-(\partial_t \Delta h_{xy})^2 +(\partial_z \Delta h_{xy})^2}$.  

As was pointed out by Milgrom \cite{Milgrom:2013iea}, any function of $t\pm z$ is an exact solution to 
Eq.~(\ref{eq:hxyeq}).
Nevertheless Eq.~(\ref{eq:hxyeq}) is problematic for gravitational wave physics for several reasons. 
First and foremost, Eq.~(\ref{eq:hxyeq}) is non-linear and consequently superpositions of
solutions propagating in opposite directions are not themselves solutions.  
This means that gravitational waves propagating in different directions 
will interact with each other, with the 
outgoing waveforms being modified by the interaction.  
This alone would suggest that 
gravitational waves in BIMOND originating from merger events should not resemble those in
GR, which is in contradiction with data from LIGO.  

Moreover, the violations 
of the superposition principle can be violent, with interactions resulting
in singularities.   
In particular, at points in which 
$2 (\partial_t \Delta h_{xy})^2 =  (\partial_z \Delta h_{xy})^2$, Eq.~(\ref{eq:hxyeq})
implies that $\partial_t^2 \Delta h_{xy}$ generically diverges.  
For example, a superposition of two plane waves initially propagating in the $\pm z$ directions
can have divergent $\partial_t^2 \Delta h_{xy}$.  
We have also numerically solved Eq.~(\ref{eq:hxyeq}) for the collision 
of two initially well separated Gaussian wave packets propagating towards each other 
at the speed of light. We have found that as soon as the tails of the Gaussian begin to 
overlap, interactions naturally lead to points in which $(\partial_t \Delta h_{xy})^2 = 2 (\partial_z \Delta h_{xy})^2$
and correspondingly, $\Delta h_{xy}$ becoming singular %
\footnote
  {In the absence of background potentials,
  similar instabilities also exist in the vector and scalar modes in above Einstein-Aether theory.
  }.
While its possible that higher order corrections to the scaling relation (\ref{eq:scalingrelation})
may ameliorate the formation of singularities, the structure of gravitational wave forms will still 
be dramatically altered by interactions with background gravitational waves, and hence will not 
resemble wave forms in GR.
These features make 
BIMOND in its present form an unsatisfactory theory of modified gravity.

\section{Discussion}%
In this paper we have argued that gravitational waves should obey two simple principles:
(i) they should propagate at the speed of light and (ii) they should obey 
linear equations in the weak-field limit.  The first condition follows from considerations 
of energy loss of cosmic rays
via gravitational Cherenkov radiation \cite{Moore:2001bv}, while the second is supported by recent 
observations of gravitational waves by LIGO.
It is striking how these two simple principles 
constrain the space of allowed MONDian theories of gravity.  
It is worth noting that gravitational waves in TeVeS also generically 
propagate at speeds different from the speed of light \cite{Sagi:2010ei}.

We note that Milgrom \cite{Milgrom:2011qt} has argued that MONDian effects can
drastically reduce gravitational Cherenkov energy losses.  
Energy loss from gravitational Cherenkov radiation is dominated by the emission of radiation with momentum
$k \sim p$, with $p$ the momentum of the cosmic ray  \cite{Moore:2001bv}.
Milgrom suggested that MONDian effects can result 
in a large \textit{effective} size of cosmic rays and that radiation emitted from different points 
will deconstructively interfere, resulting in an effective momentum 
cutoff $k_{\rm max} \ll p$ and hence 
much weaker energy loss from gravitational Cherenkov radiation.  
For \textit{steady-state} sources of radiation
this is clearly the case, since the transverse size $R$ of a steady-state classical source cuts off the spectrum of excited 
modes at $k \propto 1/R$.  However, this argument cannot be correct for non-steady-state sources
of radiation.  First, the emission of Cherenkov radiation is similar to an unstable particle decaying.
Indeed, both processes are described by the same Feynman diagrams.
Milgrom's argument would suggest that large composite particles should be stable to decay via the emission of short wavelength 
radiation.  The flaw in this argument lies in the fact that it neglects the (large) recoil 
experienced by cosmic rays when radiation is emitted.
Simply put, this argument cannot apply to gravitational Cherenkov radiation irrespective of 
any effective size induced by MONDian physics.  


\section{Acknowledgments}%
This work was supported in part by the Black Hole Initiative at Harvard University, 
which is funded by a grant from the John Templeton Foundation.  We thank Guy Moore for
helpful conversations.

\bibliographystyle{utphys}
\bibliography{refs}%
\end{document}